\documentclass[a4paper]{article}
\usepackage{Odyssey2022}
\usepackage{epsfig,amssymb,amsmath}
\usepackage[subrefformat=parens]{subfig}
\usepackage[linesnumbered,ruled,vlined,commentsnumbered]{algorithm2e}
\usepackage{comment}
\usepackage{hyperref}
\usepackage{mathtools}
\usepackage{booktabs}
\usepackage{siunitx}
\usepackage{color}
\usepackage{etoolbox}

\SetCommentSty{mycommfont}
\makeatletter
\patchcmd{\@algocf@start}% <cmd>
  {-1.5em}% <search>
  {0pt}% <replace>
  {}{}% <success><failure>
\makeatother

\usepackage{xspace}
\makeatletter
\DeclareRobustCommand\onedot{\futurelet\@let@token\@onedot}
\def\@onedot{\ifx\@let@token.\else.\null\fi\xspace}
\def\eg{\emph{e.g}\onedot} 
\def\ie{\emph{i.e}\onedot}

\makeatother

\def\equationautorefname~#1\null{(#1\null)}
\newcommand{\subfigureautorefname}{\figureautorefname}

\newcommand{\vect}[1]{\mbox{\boldmath $#1$}}
\newcommand{\abs}[1]{\left\lvert#1\right\rvert}

\newcommand{\trans}[1]{#1^\mathsf{T}}

\def\appendixautorefname~#1\null{~#1 \null}

\let\orgautoref\autoref

\renewcommand{\autoref}[1]
{
\def\figureautorefname{Fig.}
\def\subfigureautorefname{Fig.}
\orgautoref{#1}
}

\ninept

\title{Improving the Naturalness of Simulated Conversations\\for End-to-End Neural Diarization}

\makeatletter
\def\name#1{\gdef\@name{#1\\}}
\makeatother
\name{{\em Natsuo Yamashita$^1$\sthanks{{\;\;}This work has been done during internship at Hitachi, Ltd.}\quad Shota Horiguchi$^2$\quad Takeshi Homma$^2$}}
\address{$^1$The University of Tokyo \\
$^2$Hitachi, Ltd.\\
{\small \tt natsuo-yamashita@g.ecc.u-tokyo.ac.jp, \{shota.horiguchi.wk,takeshi.homma.ps\}@hitachi.com} }
\begin{document}
\maketitle

\begin{abstract}
This paper investigates a method for simulating natural conversation in the model training of end-to-end neural diarization (EEND).
Due to the lack of any annotated real conversational dataset, EEND is usually pretrained on a large-scale simulated conversational dataset first and then adapted to the target real dataset.
Simulated datasets play an essential role in the training of EEND, but as yet there has been insufficient investigation into an optimal simulation method.
We thus propose a method to simulate natural conversational speech.
In contrast to conventional methods, which simply combine the speech of multiple speakers, our method takes turn-taking into account.
We define four types of speaker transition and sequentially arrange them to simulate natural conversations.
The dataset simulated using our method was found to be statistically similar to the real dataset in terms of the silence and overlap ratios.
The experimental results on two-speaker diarization using the CALLHOME and CSJ datasets showed that the simulated dataset contributes to improving the performance of EEND.


\end{abstract}

\section{Introduction}

Speaker diarization is the task of identifying speech segments and their speakers from audio or video recordings; in other words, a task to identify ``who spoke when" \cite{park2022review}.
It is widely utilized in a variety of applications such as meeting transcription \cite{yoshioka2019meeting, yoshioka2019advances}, conversational interaction analysis \cite{kumar2020improving}, content-based audio indexing \cite{guo2016remeeting}, and conversational AI \cite{addlesee2020comprehensive}.
It also helps improve the accuracy of automatic speech recognition (ASR) in multi-speaker conversations \cite{kanda2019simultaneous}.

A typical approach for speaker diarization \cite{shum2013unsupervised,sell2014speaker,garcia2017speaker,zhang2019fully,park2020auto,li2021discriminative,landini2022bayesian} is a cascade of the following steps: speech activity detection, speaker embedding extraction, and clustering.
In general, speech activity detectors and speaker embeddings extractors are constructed using neural networks, which require a large amount of training data to obtain a good performance.
Since these modules can be trained from only single-speaker recordings, they can directly leverage existing large-scale datasets such as VoxCeleb \cite{nagrani2020voxceleb} and SITW \cite{mclaren16b_interspeech}.

Alternatively, the end-to-end approach for speaker diarization is gaining attention due to its simple architecture and promising results compared to the conventional cascaded systems \cite{fujita2019selfattention,horiguchi2020end,maiti2021endtoend,liu2021end}.
In this approach, diarization models are designed to estimate each speaker's speech activities from an input multi-speaker conversational recording. Thus, they require large-scale labeled conversational recordings for training, but unfortunately, the amount of such labeled conversational data is limited---at least compared to the single-speaker datasets \cite{carletta2007unleashing,barker2018fifth}.
To deal with the problem of limited real data, past studies on end-to-end neural diarization (EEND) \cite{fujita2019end,fujita2019selfattention,horiguchi2020end,horiguchi2021encoder,maiti2021endtoend} have utilized models that are first pretrained using simulated conversational data created from single-speaker datasets and then adapted with real data.
Such simulated conversational data can be generated with infinite variation given single-speaker recordings, and reports have shown that using the simulated data for training can improve performance compared to training with only a small amount of real data \cite{fujita2019end,fujita2019selfattention}.

Recent studies have shown that using both simulated and real recordings during pretraining improves the diarization performance \cite{liu2021end}, especially when Conformer encoders \cite{gulati2020conformer} are used as a backbone architecture.
It is important that the turn-taking property be natural in this case because Conformer, unlike Transformer \cite{vaswani2017attention} \footnote{Transformer encoders without positional encoding are usually used for EEND.}, captures temporal context, and thus using real conversational data during pretraining helps.
Here, a research question arises as to whether improving the protocol to simulate more natural conversational data will help improve the performance of EEND.

In this paper, we propose a method to simulate natural conversational mixtures to improve the performance of EEND models.
We consider the turn-taking relationship between multi-speaker utterances by introducing four utterance transition types and mixing the utterances in a sequential manner.
As our experimental results demonstrate, the proposed simulation method improves the conversational similarity between the simulated and real data, and significantly decreases the diarization error rates (DERs) of the Transformer- and Conformer-based EEND models.

\section{Related work}

\subsection{End-to-end neural diarization}
End-to-end neural diarization (EEND) is a framework to estimate multi-speaker speech activities from an input recording simultaneously.
Given $T$-length $D$-dimensional frame-wise acoustic features $\left(\vect{x}_t\mid\vect{x}_t\in\mathbb{R}^D\right)_{t=1}^T$, the EEND model $f_\mathsf{EEND}$ directly estimates the joint speech activities of all $S$ speakers for each frame in an end-to-end fashion as
\begin{align}
    \vect{p}_1,\dots,\vect{p}_T=f_\mathsf{EEND}\left(\vect{x}_1,\dots,\vect{x}_T\right),
\end{align}
where $\vect{p}_t\coloneqq\trans{\left[p_{t,1},\dots,p_{t,S}\right]}\in\left(0,1\right)^S$ is the posterior probabilities of $S$ speakers' speech activities at $t$.
The backbone architecture of $f_\mathsf{EEND}$ can be bidirectional long short-term memory \cite{fujita2019end}, Transformer \cite{fujita2019selfattention}, Conformer \cite{liu2021end}, or a time-dilated neural network \cite{maiti2021endtoend}.
During training, the posterior probabilities are optimized to minimize the following permutation-free loss:
\begin{align}
    \mathcal{L}=\frac{1}{TS}\min_{\phi\in\Phi\left(S\right)}\sum_{t=1}^{T}H\left(\vect{y}_t,P_\phi\vect{p}_t\right),\label{eq:loss}
\end{align}
where $\phi$
where $\Phi\left(S\right)$ is the set of all possible permutation of $S$ elements, $P_\phi\in\left\{0,1\right\}^{S\times S}$ is the permutation matrix corresponding to the permutation $\phi$, $\vect{y}_t\coloneqq\trans{\left[y_{t,1},\dots,y_{t,S}\right]}\in\left\{0,1\right\}^S$ is the ground-truth speaker activities, and $H\left(\cdot,\cdot\right)$ is the binary cross entropy, defined as
\begin{align}
    H\left(\vect{y}_t,\vect{p}_t\right)\coloneqq\sum_{s=1}^{S}\left\{-y_{t,s}\log p_{t,s}-\left(1-y_{t,s}\right)\log\left(1-p_{t,s}\right)\right\}.
    \label{eq:bce}
\end{align}

Since the training objective in \autoref{eq:loss} includes the ground-truth labels $\vect{y}_t$, the training of EEND requires data with frame-wise annotations, with the exception of one study that utilized unsupervised domain adaptation of EEND \cite{takashima2021semisupervised}.
The amount of real data in the target domain is sometimes limited, so most of the previous studies on EEND have utilized simulated data \cite{fujita2019end,fujita2019selfattention,horiguchi2020end,horiguchi2021encoder,maiti2021endtoend}, where the EEND models are first pretrained using large-scale simulated multi-speaker datasets and then adapted to the real data in the target domain.
Most of the studies used the simulation method proposed in the initial EEND paper \cite{fujita2019end,fujita2019selfattention} (see \autoref{sec:conventional_simulation_method} for the detailed protocol), but there has been no investigation into the validity of the simulation method.
A recent study \cite{liu2021end} has shown that using both simulated and real data during pretraining helps improve the diarization performance.
This improvement is greater when Conformer encoders are used as the backbone architecture, as they can capture temporal contexts more effectively than Transformer encoders\footnote{Most EEND studies have used Transformer encoders without positional encodings; thus, temporal context is completely ignored.}.
This result suggests that the temporal context, or \textit{turn-taking} behavior, of the data used for pretraining should be natural, especially when Conformer encoders are used, as was discussed in \cite{liu2021end}.
In our work, we aim to provide a novel method for simulating conversational speech with natural turn-taking.

\subsection{Simulated mixtures in speech processing}
\label{sec:exiting_protocol}
The simulation of large-scale multi-speaker datasets is an important step in training data-hungry neural-network-based speech processing models, not just speaker diarization models.
In the field of speech separation, WSJ0-2mix \cite{hershey2016deep} is the first and still the most commonly used benchmark dataset.
Each mixture in WSJ0-2mix consists of near-fully overlapped utterances of two speakers, each of which is from the Wall Street Journal (WSJ0) corpus \cite{paul1992design}.
The WHAM! \cite{Wichern2019WHAM} and WHAMR! \cite{Maciejewski2020WHAMR} datasets are extensions of WSJ0-2mix to the noisy and reverberant conditions, respectively.
Libri2Mix \cite{cosentino2020librimix} has a similar overlap ratio to WSJ0-2mix, but the LibriSpeech dataset is used as the source dataset instead of the WSJ0 corpus.
To evaluate speech separation performance under various overlap ratio conditions, LibriCSS \cite{chen2020continuous} and SparseLibri2mix \cite{cosentino2020librimix} have also been proposed, both of which are created from the LibriSpeech dataset.
LibriCSS has an overlap ratio of \si{0}--\SI{40}{\percent}, and SparseLibri2mix has an overlap ratio of \si{0}--\SI{100}{\percent}.

Simulated mixtures have also been used in ASR studies.
The most commonly used type of simulated data consists of mixtures that each contain one utterance per speaker \cite{seki2018purely,kanda2020serialized,tripathi2020end}.
Such data is created by adding utterances of multiple speakers with random delays.
One study evaluated the mixtures with various turn-taking configurations, rather than simply adding utterances with delays \cite{yoshioka2018multi}.
Other studies have used long-form simulated mixtures in which each speaker utters multiple times for training. 
In the research on diarization, while cascaded approaches only require single-speaker recordings to train a speaker embedding extractor, end-to-end approaches require multi-speaker recordings for the model training \cite{von2019all,kinoshita2020tackling,chang2021hypothesis}.

As described above, simulated mixtures have been widely utilized in speech processing research.
However, the quality of the simulation, \ie, how realistic the simulated multi-speaker conversation is, has rarely been investigated.
In the case of speech separation and ASR, the evaluation data is also often simulation data, so the realism of the simulation data may not be as important.
On the other hand, in the case of speaker diarization, where evaluation with long-form real recordings is common, it is important to know how realistic the simulation data is.
In this paper, we clarify the difference in the performance of EEND depending on the simulation data used for training.

\subsection{Conversation analysis}
For better understanding and to establish a simulation protocol for speech diarization, we referred to research on conversation analysis.
In the field of psychology, turn-taking is regarded as one of the fundamental mechanisms in conversation because conversation is a type of social interaction \cite{levinson2015timing, sacks1978simplest}.
Prior studies have analyzed the temporal patterning of turn-taking and categorized the relations between multi-speaker utterances into patterns such as gap, pause, and overlap \cite{heldner2010pauses} or continuation, interrupt, and turn change \cite{ten2004turn}.
Inspired by their work, we take human turn-taking behaviors into account to create training data for EEND.
On the basis of these earlier studies on categorization, we introduce four transition types between utterances, as described in \autoref{sec:transition_type}.

\section{Conventional simulation method: Concat-and-sum approach}
\label{sec:conventional_simulation_method}
The mixture simulation protocol used in the previous EEND studies \cite{fujita2019end, fujita2019selfattention} can be referred to as the \textit{concat-and-sum} approach.
In order to create an $S$-speaker mixture, a long recording $\vect{x}^{(s)}$ for each speaker $s\in\left\{1,\dots,S\right\}$ is first prepared, where $\vect{x}^{(s)}=(x^{(s)}_1,\dots,x^{(s)}_{T_s})\in\mathbb{R}^{T_s}$ is the $T_s$-length discrete waveform of $s$-th speaker's speech, \ie, $\abs{\vect{x}^{(s)}}=T_s$.
Given $N_s$ utterances $\left\{\vect{u}^{(s)}_{n}\right\}_{n=1}^{N_s}$ of the $s$-th speaker, each of which is also the discrete waveform, they are concatenated with a silence between them as
\begin{align}
    \vect{x}^{(s)}=\vect{u}^{(s)}_{1}\oplus\vect{0}^{(s)}_{1}\oplus\vect{u}^{(s)}_{2}\oplus\vect{0}^{(s)}_{2}\oplus\dots\oplus\vect{0}^{(s)}_{N_{s-1}}\oplus\vect{u}^{(s)}_{N_s},
\end{align}
where $\oplus$ is the tuple concatenation operator and $\vect{0}^{(s)}_{k}$ for $1\leq k\leq N_{s-1}$ is the all-zero tuple, whose length $\delta$ is sampled from the exponential distribution of the expected value $\beta$.
Then, a random room impulse response $\vect{i}^{(s)}$ is convolved to each speaker's long recording, as
\begin{align}
    \vect{x}^{(s)}\leftarrow\vect{x}^{(s)}*\vect{i}^{(s)}.
\end{align}
Finally, all the long recordings and the $L$-length noise signal $\vect{x}_\text{noise}$ are concatenated to be an $S$-speaker mixture $\vect{x}$, as
\begin{align}
    \vect{x}&=\vect{x}_\text{noise}+\sum_{s=1}^{S}\vect{x}^{(s)}\oplus(\underbrace{0,\dots,0}_{L-T_s}),\label{eq:concat}\\
    L&=\max_{s\in\left\{1,\dots,S\right\}}T_s.
\end{align}

\begin{figure}[t]
\centering
    \subfloat[CALLHOME1]{%
        \includegraphics[width=0.49\linewidth]{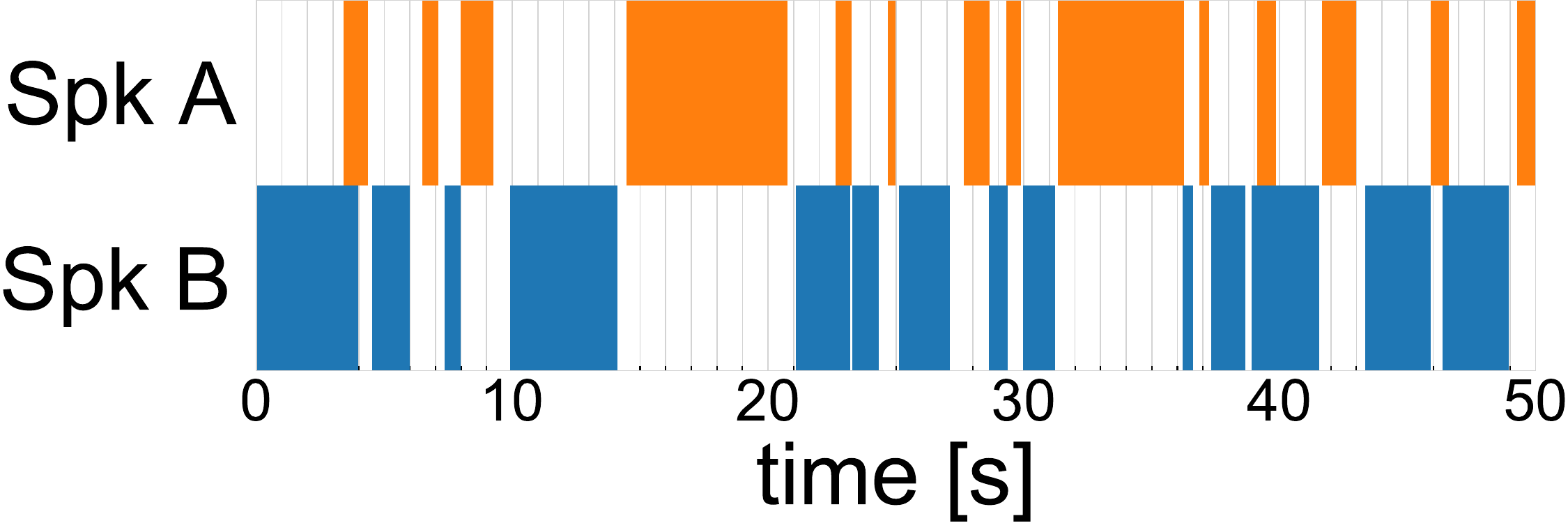}%
        \label{fig:visualize_callhome}%
    }
    \hfill
    \subfloat[CSJ]{%
        \includegraphics[width=0.49\linewidth]{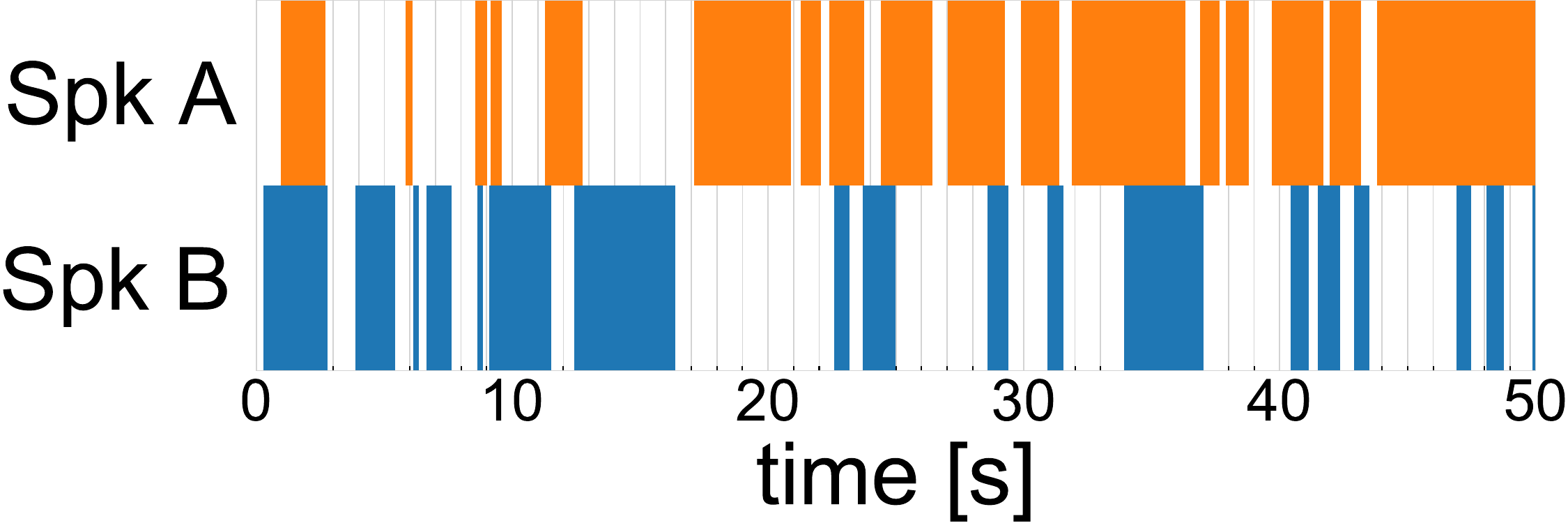}%
        \label{fig:visualize_csj}%
    }
  \\
  \subfloat[Concat-and-sum ($\beta=2$)]{%
    \includegraphics[width=0.49\linewidth]{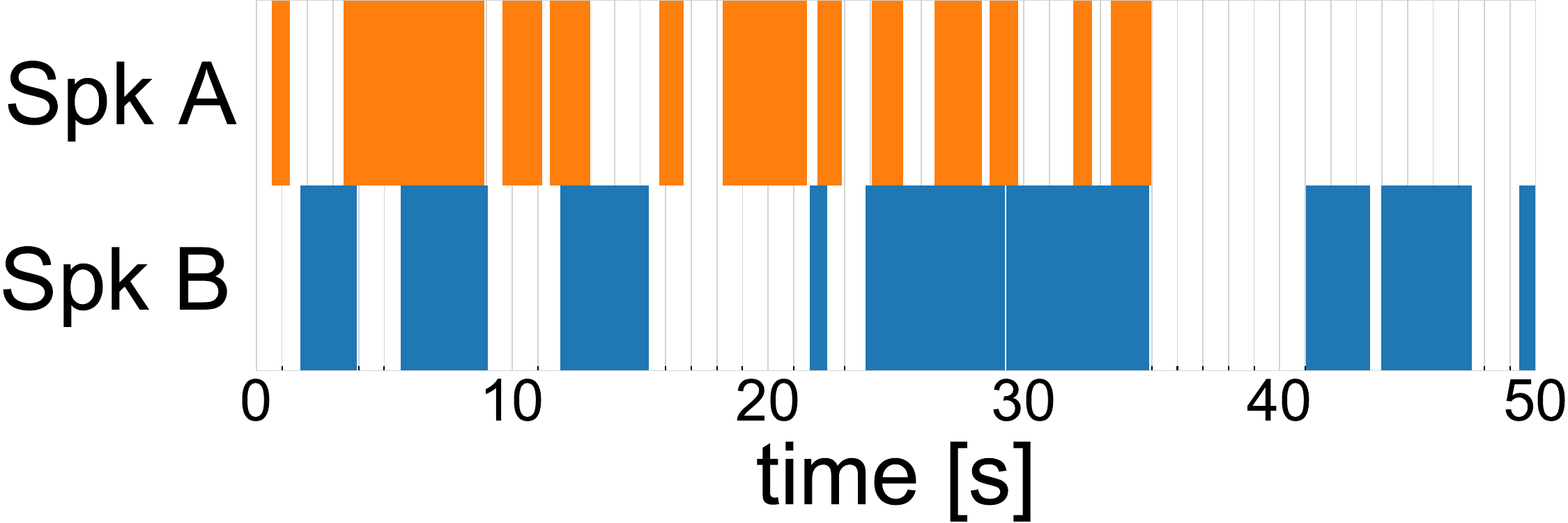}%
    \label{fig:visualize_beta2}%
  }
  \hfill
  \subfloat[Concat-and-sum ($\beta=7$)] {%
    \includegraphics[width=0.49\linewidth]{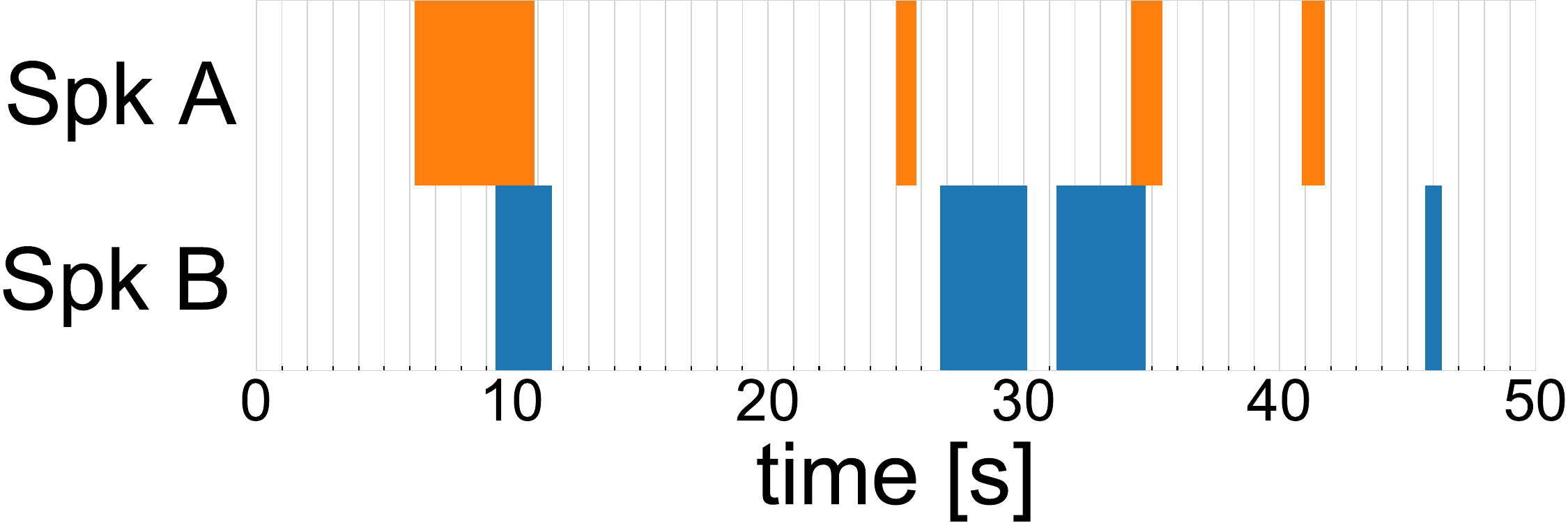}
    \label{fig:visualize_beta7}
  }
   \\
  \subfloat[Ours (Random selection)]{%
    \includegraphics[width=0.49\linewidth]{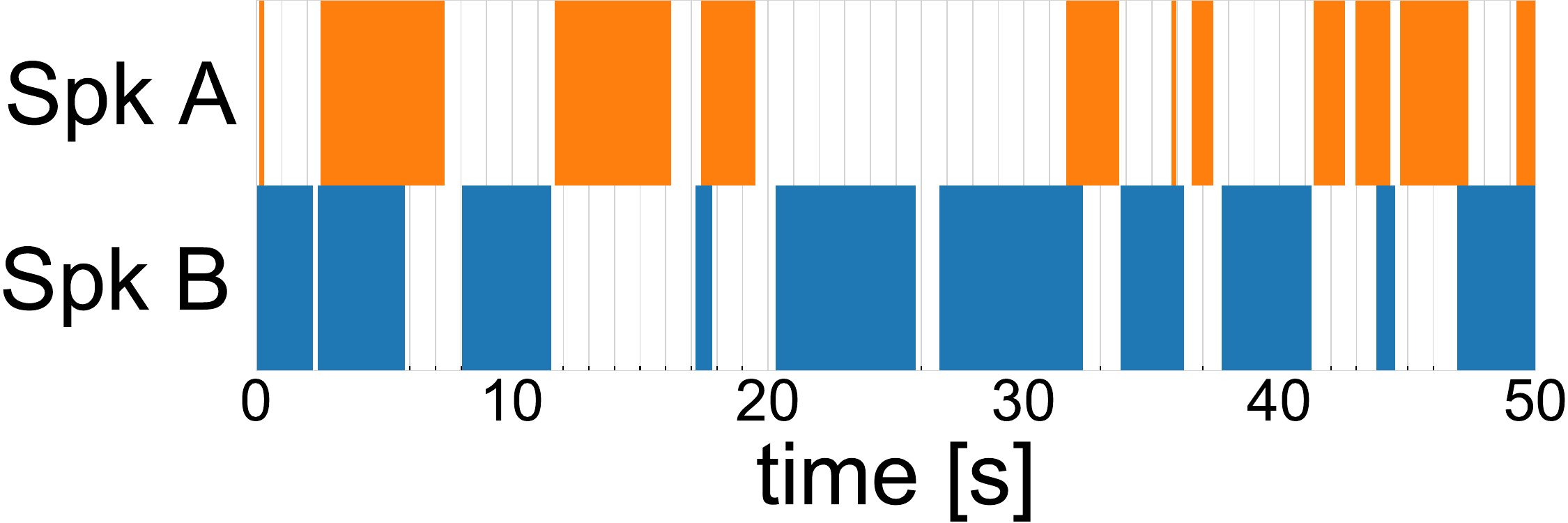}%
    \label{fig:visualize_random}%
  }
  \hfill
  \subfloat[Ours (Markov selection)]{%
    \includegraphics[width=0.49\linewidth]{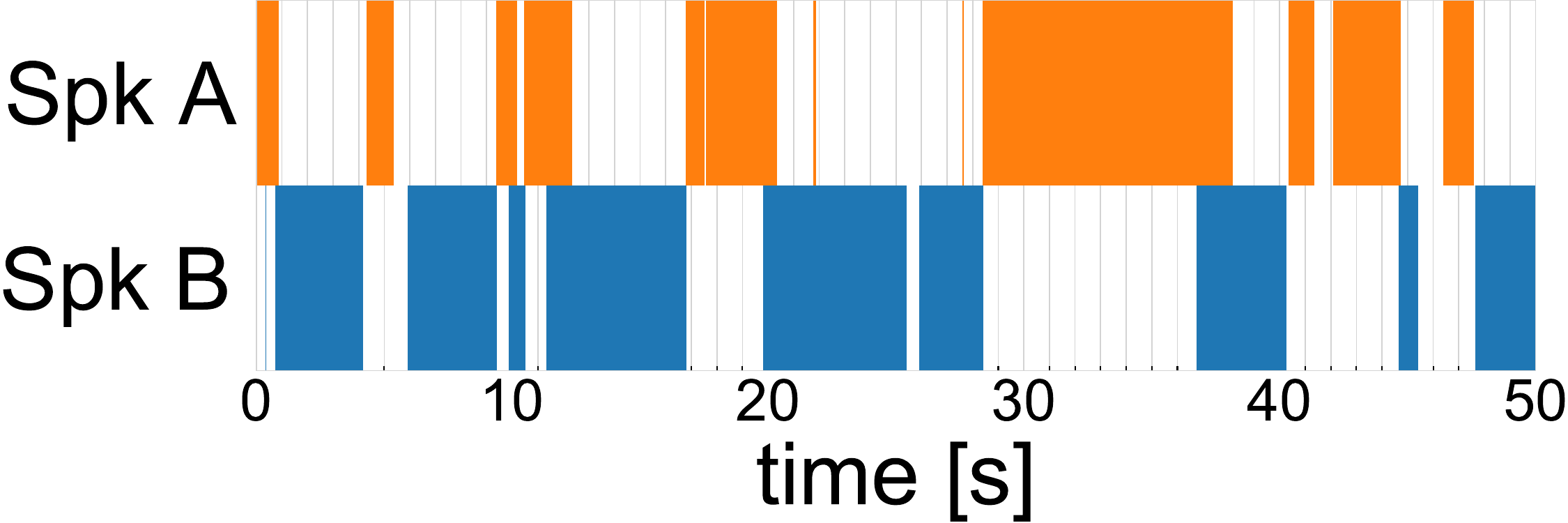}%
    \label{fig:visualize_markov}%
  }
\caption{\textit{Visualization of two-speaker conversations of real datasets (top), simulated datasets by the conventional method (middle), and simulated datasets by the proposed method (bottom).}}
\label{fig:visualize}
\end{figure}

In the simulation protocol above, relations between utterances of different speakers are not considered. As a result, speakers appearing in the simulated mixtures do not speak alternately (as in a real conversation shown in Figs.~\ref{fig:visualize_callhome} and \ref{fig:visualize_csj}), and there are unnatural silence intervals and overlapping speech, as shown in Figs.~\ref{fig:visualize_beta2} and \ref{fig:visualize_beta7}.

Indeed, it is possible to control the statistics of the simulated mixtures, \eg, the overlap ratio, by varying the value of $\beta$ in the exponential distribution.
However, reducing the overlap ratio by increasing the value of $\beta$ also makes the duration of silence longer; thus, it is difficult to make the simulated mixture closer to a natural conversation by simply adjusting the value of $\beta$.
Moreover, concatenating a zero vector in \autoref{eq:concat} to align the lengths of the long recordings causes unnaturalness of the conversation; various speakers speak at the beginning of the mixtures while only some of them speak at the end.
In contrast, the proposed simulation protocol can generate more natural conversations by ordering each speaker's utterances to follow the statistics calculated from real conversational data.

\section{Proposed simulation method}
\label{sec:ours}
In the proposed simulation method, utterances of multiple speakers are sequentially arranged as in a real conversation.
This makes it possible to avoid unnatural conversational patterns, silence, and overlapping speech, as shown in \autoref{fig:visualize_random} and \autoref{fig:visualize_markov}.
The following subsections show how the utterance transition types are defined and arranged to simulate a conversation.

\subsection{Transition types between utterances}\label{sec:transition_type}

\label{sec:patterns}

\begin{figure}[tb]
  \centering
  \subfloat[Turn-hold]{
    \includegraphics[width=0.475\linewidth]{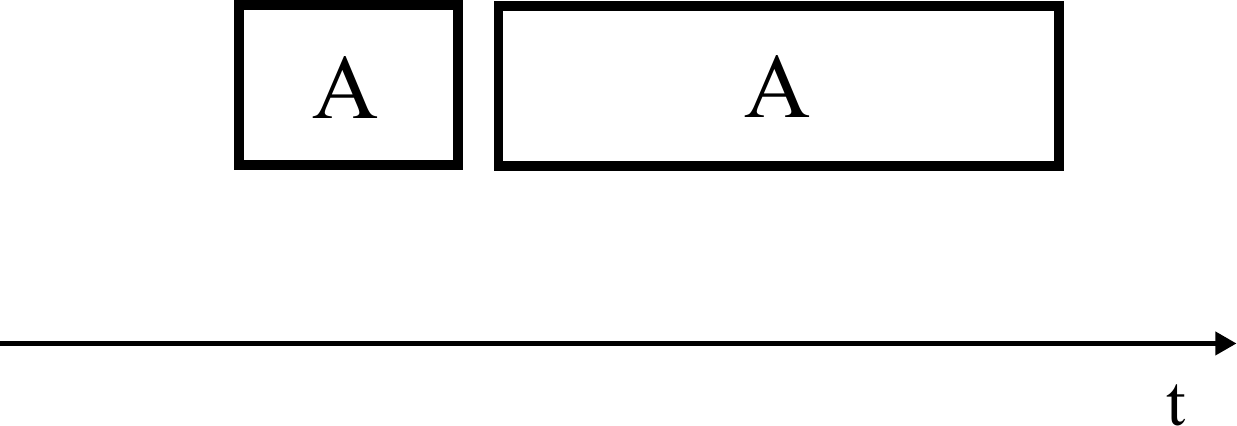}
    \label{fig:turn_hold}
  }
  \hfill
  \subfloat[Turn-switch]{
    \includegraphics[width=0.475\linewidth]{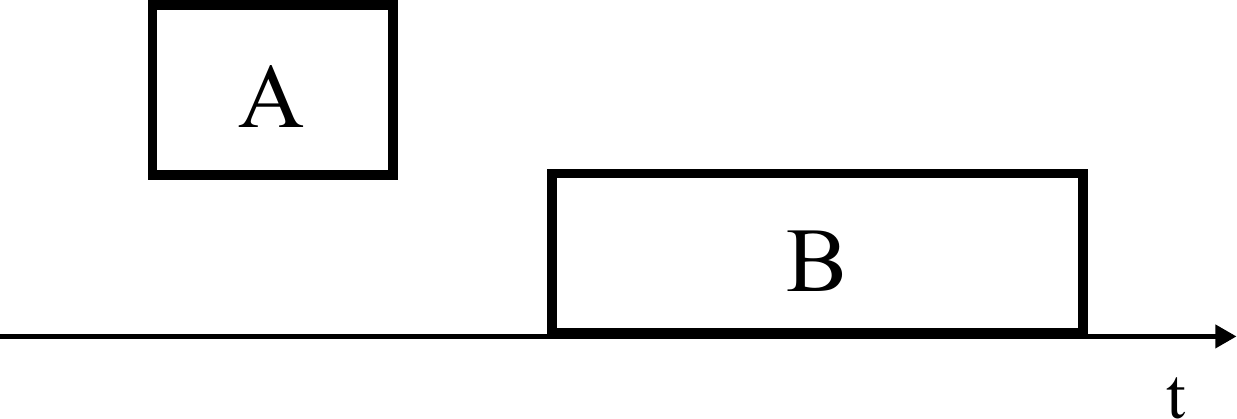}
    \label{fig:turn_switch}
  }
  \\
  \subfloat[Interruption]{
    \includegraphics[width=0.475\linewidth]{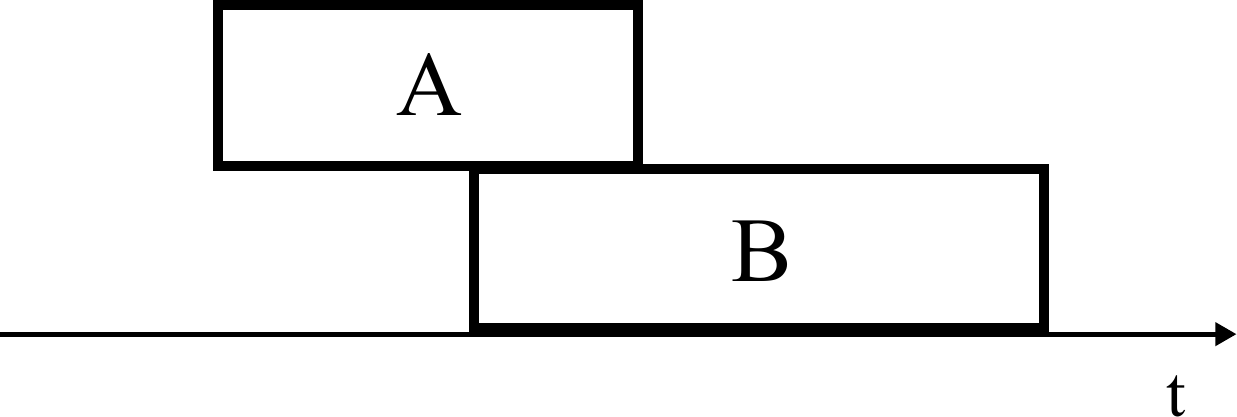}
    \label{fig:interruption}
  }
  \hfill
  \subfloat[Backchannel]{ \includegraphics[width=0.475\linewidth]{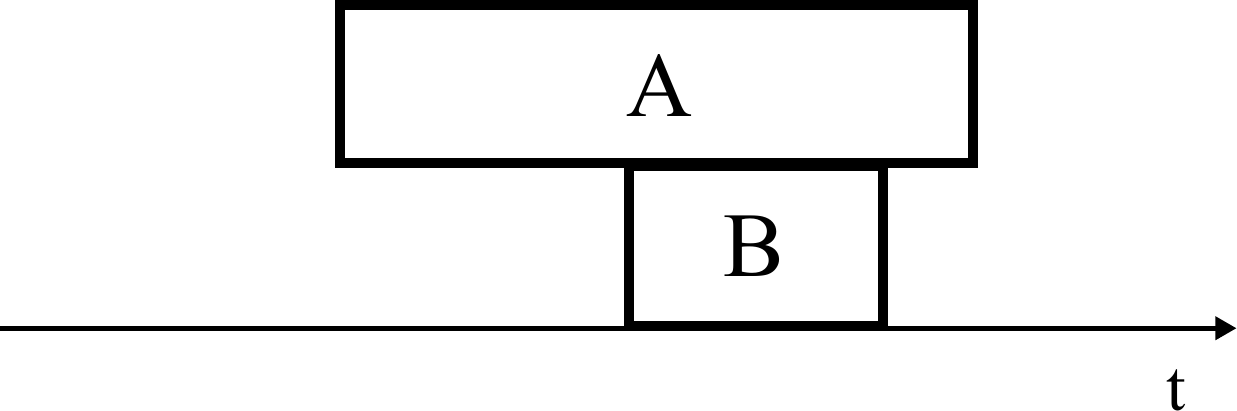}
    \label{fig:backchannel}
  }
  \caption{\textit{Four utterance transition cases in two-speaker scenario.}}
  \label{fig:patterns}
\end{figure}

In the proposed simulation protocol, utterances are arranged in an autoregressive manner.
One possible approach to this is to decide the next speaker based on the current speaker, as in VBx \cite{landini2022bayesian}.
However, VBx only considers which speaker to transition to next, not how to transition.
In other words, it does not take silence or overlap between utterances into account.
Our method, in contrast, also takes into account how to transition to the next speaker by introducing variations determined between a pair of successive utterances.

Assuming that at most two speakers speak at the same time, there are four possible utterance transition cases.
Let $\vect{u}_\text{prev}$ be the utterance with the latest end time among the utterances arranged so far, and $\vect{u}_\text{next}$ be the next utterance to be arranged.
The four transition patterns of $\vect{u}_\text{prev}$ and $\vect{u}_\text{next}$ are shown in \autoref{fig:patterns} and can be described as follows.

\begin{itemize}
    \item \textbf{Turn-hold (TH)} (\autoref{fig:turn_hold}): Two utterances from the same speaker with inter-turn silence (pause) between them. We assume that the length of pause follows the exponential distribution with the expected value $\beta_\text{TH}$.
    \item \textbf{Turn-switch (TS)} (\autoref{fig:turn_switch}): Two utterances from different speakers with intra-turn silence (gap) between them. We assume that the length of pause follows the exponential distribution with the expected value $\beta_\text{TS}$.
    \item \textbf{Interruption (IR)} (\autoref{fig:interruption}): Two utterances by different speakers that partially overlap each other.
    \item \textbf{Backchannel (BC)} (\autoref{fig:backchannel}): Two utterances by different speakers, one of which is fully overlapped with the other.
    This is also known as \textit{aizuchi}.
\end{itemize}

For turn-hold and turn-switch, we assume that the duration of each silence $\delta$ between $\vect{u}_\text{prev}$ and $\vect{u}_\text{next}$ follows an exponential distribution, \ie,
\begin{align}
    \delta\sim f\left(\delta;\beta\right)\coloneqq\frac{1}{\beta}\exp\left(-\frac{\delta}{\beta}\right).\label{eq:silence}
\end{align}
We use the different expected values for turn-hold and turn-switch, each of which are denoted as $\beta_\text{TH}$ and $\beta_\text{TS}$, respectively.

For interruption and backchannel, on the other hand, the duration of overlap cannot be determined in the same way, since the randomly determined overlap duration may be larger than $\abs{\vect{u}_\text{prev}}$ and $\abs{\vect{u}_\text{next}}$.
Therefore, we instead determine the ratio of the overlap $\rho$.
Let $\vect{u}'_\text{prev}$ be the last part of $\vect{u}_\text{prev}$, which is not overlapped with all the utterances before $\vect{u}_\text{next}$; then, the overlap duration is calculated as
\begin{align}
    \delta=\rho\cdot\min\left(\abs{\vect{u}'_\text{prev}},\abs{\vect{u}_\text{next}}\right),\label{eq:overlap_1}
\end{align}
where $\rho$ is sampled from the truncated exponential distribution whose probability density function is defined as
\begin{align}
    \rho\sim g\left(\rho;\beta\right)\coloneqq\begin{cases}\frac{
    f\left(\rho;\beta\right)}{\int_{\epsilon}^{1-\epsilon}f\left(\rho';\beta\right)d\rho'}&(\epsilon\leq\rho\leq 1-\epsilon)\\
    0 &\text{(otherwise)}\label{eq:overlap_2}
    \end{cases},
\end{align}
where $\epsilon\ll1$ is set to 0.03 in this paper.
In this case too, we use the different values $\beta$ for interruption and backchannel, each of which are denoted as $\beta_\text{IR}$ and $\beta_\text{BC}$, respectively.
Note that the starting time of $\vect{u}_\text{next}$ is uniformly determined in the case of backchannel.

After the arrangement of $\vect{u}_\text{next}$, $\vect{u}_\text{prev}$ is updated for the next step.
In the case of turn-hold, turn-switch, and interruption, $\vect{u}_\text{prev}$ is updated as
\begin{align}
    \vect{u}_\text{prev}\leftarrow\vect{u}_\text{next},
\end{align}
while it is not updated in the case of backchannel.

\subsection{Sequential determination of transition types}\label{sec:sequential_determination}
\label{sec:selection}
In our mixture simulation protocol, multi-speaker utterances are mixed by sequentially determining the next speaker and transition type.
We consider two types of transition type generation: \textit{random selection} and \textit{Markov selection}.

\textbf{Random selection}:
The random selection assumes that the transition type at each transition is independent of the others, \ie, the transition type is randomly selected according to the probability $P_\text{ind}$:
\begin{align}
    P_\text{ind}=\left[p_\text{TH},p_\text{TS},p_\text{IR},p_\text{BC}\right],
\end{align}
where $p_\text{X}$ for $X\in\left\{\text{TH},\text{TS},\text{IR},\text{BC}\right\}$ is the probability of occurrence of each transition type.

\textbf{Markov selection}:
The Markov selection assumes that the transition types follow a simple Markov process, \ie, the probability of each transition type depends on the previous transition type.
The transition matrix $P\in\left[0,1\right]^{4\times4}$ is denoted as
\begin{align}
    P_\text{Markov}=\begin{bmatrix}
        p_\text{TH$\to$TH}&p_\text{TS$\to$TH}&p_\text{IR$\to$TH}&p_\text{BC$\to$TH}\\
        p_\text{TH$\to$TS}&p_\text{TS$\to$TS}&p_\text{IR$\to$TS}&p_\text{BC$\to$TS}\\
        p_\text{TH$\to$IR}&p_\text{TS$\to$IR}&p_\text{IR$\to$IR}&p_\text{BC$\to$IR}\\
        p_\text{TH$\to$BC}&p_\text{TS$\to$BC}&p_\text{IR$\to$BC}&p_\text{BC$\to$BC}\\
    \end{bmatrix},
    \label{equ:transition_matrix}
\end{align}
where $p_{X\to Y}$ for $X,Y\in\left\{\text{TH},\text{TS},\text{IR},\text{BC}\right\}$ is the probability that the next state will be $Y$ given the current state $X$.

The parameters $P_\text{ind}$ and $P_\text{Markov}$ decide the naturalness of the simulated mixtures.
In this paper, we use these parameters extracted from the real conversational data.

\subsection{Overall simulation protocol}
\begin{algorithm}[t]
	\SetAlgoLined
	\DontPrintSemicolon
	\caption{Mixture simulation.}
	\label{alg:mixture_simulation}
	\SetAlgoVlined
	\SetKwInOut{Input}{Input}
	\SetKw{In}{in}
	\Input{
	    {{$\mathcal{S,N,I,R}$} \tcp*{Set of speakers, noises, RIRs, and SNRs}}\\
	    {{$\mathcal{U}_1,\dots,\mathcal{U}_{\abs{\mathcal{S}}}$} \tcp*{Set of utterances for each speaker}}
	    {{$N_\text{spk},N_\text{utt}$} \tcp*{No. of speakers / no. of utterances per mixture}}
	}
	\SetKwInOut{Output}{Output}
	\Output{$\vect{x}$\tcp*{Mixture}}
	\BlankLine
	
	Sample a set of $N_\text{spk}$ speakers $\mathcal{S'}$ from $\mathcal{S}$\\
    Sample $\vect{i}^{(s)}$ for $s\in\mathcal{S}'$ from $\mathcal{I}$\tcp*{RIR}
	Sample the first speaker $s_1$ from $\mathcal{S}'$\\
	Initialize $\vect{x}$ with $\vect{u}*\vect{i}^{(s_1)}$, where $\vect{u}\sim U_{s_1}$\\
	\For{$t\leftarrow2$ \KwTo $N_\text{\rm utt}$}{
        Decide the next utterance transition type $Z_t$\\
        \uIf(\tcp*[f]{Turn-hold}){$Z_t=\text{\rm TH}$}{
            $s_t=s_{t-1}$\\
            Sample $\vect{u}$ from $\mathcal{U}_{s_t}$\\
            Prepare a silence signal $\vect{0}_t$ using \autoref{eq:silence} \\
            $\vect{x}\leftarrow\vect{x}\oplus\vect{0}_t\oplus\vect{u}*\vect{i}^{(s_t)}$\\
        }
        \uElseIf(\tcp*[f]{Turn-switch}){$Z_t=\text{\rm TS}$}{
            Sample $s_t$ from $\mathcal{S'}\setminus\left\{s_{t-1}\right\}$\\
            Sample $\vect{u}$ from $\mathcal{U}_{s_t}$\\
            Prepare a silence signal $\vect{0}_t$ using \autoref{eq:silence} \\
            $\vect{x}\leftarrow\vect{x}\oplus\vect{0}_t\oplus\vect{u}*\vect{i}^{(s_t)}$\\
        }
        \uElseIf(\tcp*[f]{Interruption}){$Z_t=\text{\rm IR}$}{
            Sample $s_t$ from $\mathcal{S'}\setminus\left\{s_{t-1}\right\}$\\
            Sample $\vect{u}$ from $\mathcal{U}_{s_t}$\\
            $\vect{u}\leftarrow\vect{u}*\vect{i}^{(s_t)}$\\
            Decide the overlap duration $\delta$ using \autoref{eq:overlap_1}\autoref{eq:overlap_2}\\
            $\vect{x}\leftarrow\vect{x}_{1:\abs{\vect{x}}-\delta}\oplus\left(\vect{x}_{\abs{\vect{x}}-\delta:\abs{\vect{x}}}+\vect{u}_{1:\delta}\right)\oplus\vect{u}_{\delta:\abs{\vect{u}}}$
        }
        \ElseIf(\tcp*[f]{Backchannel}){$Z_t=\text{\rm BC}$}{
            Sample $s_t$ from $\mathcal{S'}\setminus\left\{s_{t-1}\right\}$\\
            Sample $\vect{u}$ from $\mathcal{U}_{s_t}$\\
            $\vect{u}\leftarrow\vect{u}*\vect{i}^{(s_t)}$\\
            Decide the overlap duration $\delta$ using \autoref{eq:overlap_1}\autoref{eq:overlap_2}\\
            Decide the start time $\tau$ of $\vect{u}$\\
            $\vect{x}\leftarrow\vect{x}_{1:\tau}\oplus\left(\vect{x}_{\tau+\abs{\vect{u}}}+\vect{u}\right)\oplus\vect{x}_{\tau+\abs{\vect{u}}:\abs{\vect{x}}}$
        }
	}
	Sample $\vect{n}$ from $\mathcal{N}$\tcp*{Background noise}
	Sample $r$ from $\mathcal{R}$\tcp*{SNR}
	Determine a mixing scale $p$ from $r,\vect{x},$ and $\vect{n}$\\
	$\vect{n}'\leftarrow$ repeat $\vect{n}$ until reaching the length of $\vect{x}$\\
	$\vect{x}\leftarrow\vect{x}+p\cdot\vect{n}'$\\
\end{algorithm}

The algorithm of the proposed simulation protocol is presented in \autoref{alg:mixture_simulation}.
As inputs, set of speakers $\mathcal{U}$, noises $\mathcal{N}$, room impulse responses (RIRs) $\mathcal{I}$, signal-to-noise ratios $\mathcal{R}$, and utterances for each speaker $\mathcal{U}_1,\dots,\mathcal{U}_{\abs{\mathcal{S}}}$ are given.
The numbers of speakers $N_\text{spk}$ and utterances $N_\text{utt}$ within a mixture are also given as inputs.

First, the set of $N_\text{spk}$ speakers and corresponding RIRs are sampled (L1--L2).
The speakers and their utterances are determined sequentially and arranged on the basis of the utterance transition types described in \autoref{sec:transition_type} and \autoref{sec:sequential_determination} to construct a long-form audio (L3--L29).
Finally, a noise is added with a randomly determined SNR to obtain a final simulated mixture (L30--34).

\section{Experiments}

\subsection{Data}
\subsubsection{Real datasets}
We utilized the two-speaker portion of the CALLHOME dataset \cite{callhome} for the experiment.
Following the setting in previous studies \cite{fujita2019selfattention,horiguchi2020end}, we used the split provided in the Kaldi recipe\footnote{\url{https://github.com/kaldi-asr/kaldi/tree/master/egs/callhome_diarization/v2}} to create an adaptation set consisting of 155 recordings (CALLHOME1) and a test set consisting of 148 recordings (CALLHOME2).
We also used 54 dialogue recordings in the Corpus of Spontaneous Japanese (CSJ) for evaluation.

\subsubsection{Simulated mixtures}
The sets of utterances used for the simulation were extracted from the Switchboard-2 (Phase I, II, III), Switchboard Cellular (Part1, Part2), and NIST Speaker Recognition Evaluation (2004, 2005, 2006, 2008) datasets.
All recordings are telephone speech sampled at \SI{8}{\kHz}.
There are 6,381 speakers in total, and we used 5,743 speakers among them to create the training sets following previous studies \cite{fujita2019end,fujita2019selfattention,horiguchi2020end,horiguchi2021encoder,liu2021end}.
We extracted utterances from these datasets by using the time-delay-neural-network-based speech activity detector\footnote{\url{https://github.com/kaldi-asr/kaldi/tree/master/egs/aspire/s5}}.
While most previous studies filtered out detected utterances of less than \SI{1.5}{\second}, one study reported that utilizing these short utterances helps improve diarization performance.
Therefore, in this paper we examined using utterances only \SI{1.5}{\second} or longer and using all the detected utterances.

With the set of utterances, we created two-speaker simulated conversational datasets.
For the conventional simulation method, we used $\beta=2$ following previous studies \cite{fujita2019end,fujita2019selfattention,horiguchi2020end,horiguchi2021encoder}.
We also used $\beta=7$ to obtain a similar overlap ratio to CALLHOME1.
For the proposed simulation method, we determine the parameters based on those extracted from CALLHOME1, as shown in \autoref{tbl:mixing_parameters}.
Each simulated dataset has 100,000 mixtures and the number of utterances within each mixture was adjusted so that the total duration of each dataset was roughly the same (see \autoref{tab:data_similarity}).

\begin{table}[t]
    \centering
    \caption{Parameters extracted from CALLHOME1.}
    \label{tbl:mixing_parameters}
    \begin{tabular}{cc}
        \toprule
        Parameter&Value\\\midrule
        $\left[\beta_\text{TH},\beta_\text{TS},\beta_\text{IR},\beta_\text{BC}\right]$ &  $\left[0.57,0.40,0.10,0.44\right]$\\\midrule
        $P_\text{ind}$ &$\left[0.15,0.31,0.44,0.10\right]$\\\midrule
        $P_\text{Markov}$&$\begin{bmatrix}
        0.26&0.11&0.09&0.31\\
        0.23&0.38&0.29&0.29\\
        0.27&0.45&0.53&0.31\\
        0.24&0.06&0.09&0.09\\
    \end{bmatrix}$\\
        \bottomrule
    \end{tabular}
\end{table}

The set of background noises consists of 37 recordings annotated as background noises in the MUSAN corpus \cite{snyder2015musan}.
The Simulated Room Impulse Response Database \cite{ko2017study} was used as the set of RIRs for all the simulation methods.
The SNR values were sampled from $\{5,10,15,20\}$ \si{\dB}.

\subsection{Models}
We used attractor-based EEND (EEND-EDA) \cite{horiguchi2020end,horiguchi2021encoder} for the experiments.
As the backbone architecture, we used four-stacked Transformer encoders \cite{vaswani2017attention} or Conformer encoders \cite{gulati2020conformer}.
For both models, the number of attention units was set to 256, and that of attention heads to four.
The dimensionality of each position-wise feed-forward layer in the Transformer and Conformer encoders was set to 1,024 and 256, respectively.
Note that the purpose of this research is to investigate the optimal mixture simulation method; thus, we do not utilize updates introduced in the original Conformer-based EEND paper \cite{liu2021end} such as SpecAugment \cite{park2019specaugment} or convolutional subsampling.

Each model was pretrained using simulated data for 100 epochs using the Adam optimizer \cite{kingma2015adam} with a Noam scheduler \cite{vaswani2017attention} and then adapted to CALLHOME1 for another 100 epochs using Adam with a fixed learning rate of $1\times10^{-5}$.
Diarization performance was evaluated using the CALLHOME2 and CSJ datasets.

\subsection{Metrics}
We used the standard silence and overlap ratio as metrics to investigate the statistics of simulated and real conversation data, each of which are defined as
\begin{align}
    \mathrm{SilenceRatio}&=\frac{T_{\#\textit{Spk}=0}}{T_{\#\textit{Spk}=0}+T_{\#\textit{Spk}\geq1}},\\
    \mathrm{OverlapRatio}&=\frac{T_{\#\textit{Spk}\geq2}}{T_{\#\textit{Spk}\geq1}},
\end{align}
where $T_{\#\textit{Spk}=0}$, $T_{\#\textit{Spk}\geq1}$, and $T_{\#\textit{Spk}\geq2}$ denote the total duration where no speaker is active, at least one speaker is active, and multiple speakers are active, respectively.

We also utilized the similarity metric based on the earth movers' distance (EMD) \cite{liu2021end} to evaluate the distribution of the duration of overlap and silence.
Let $U$ and $V$ be the cumulative distribution functions of the overlap (or silence) duration of the datasets to be compared.
The EMD between $U$ and $V$ is determined as
\begin{align}
    \mathrm{EMD}\left(U,V\right)=\frac{\sum_{i=1}^m\sum_{j=1}^n{f_{ij}d_{ij}}}{\sum_{i=1}^m{f_{ij}}},
\end{align}
where $f_{ij}$ is the optimal flow that minimize the overall moving cost, $d_{ij}$ is the $L_1$ distance between the $i$-th element of $U$ and the $j$-th element of $V$, and $m$ and $n$ are the number of elements in $U$ and $V$, respectively.
The silence similarity or overlap similarity is determined as
\begin{align}
    \text{sim}\left(U, V\right)=\exp\left(-\gamma \mathrm{EMD}\left(U, V\right)\right),
    \label{equ:similarity}
\end{align}
where $\gamma$ is the scaling parameter, which is set to $0.001$ in this paper.

To evaluate the diarization performance, we used the diarization error rate (DER).
Following the previous studies \cite{fujita2019end,fujita2019selfattention,horiguchi2020end,horiguchi2021encoder}, we allowed \SI{0.25}{\second} of collar at the boundary of each speech segment.

\section{Results}
\subsection{Statistics of real and simulated data}

\begin{table*}[t]
    \caption{\textit{Data statistics of silence and overlap ratio, and silence/overlap similarities between real and simulated datasets. Parameters used for the proposed simulation method were extracted from CALLHOME1.}}
    \label{tab:data_similarity}
    \centering
    \resizebox{\linewidth}{!}{%
    \begin{tabular}{@{}lrrrrrrrrrr@{}}
        \toprule
        &  &  &  &  & \multicolumn{2}{c}{CALLHOME1}  & \multicolumn{2}{c}{CALLHOME2}& \multicolumn{2}{c}{CSJ} \\\cmidrule(lr){6-7}\cmidrule(lr){8-9}\cmidrule(l){10-11}
        & Utterance & Total & Silence & Overlap & Silence & Overlap & Silence & Overlap & Silence & Overlap \\
        Dataset& length & duration & ratio & ratio & similarity & similarity & similarity & similarity & similarity & similarity \\\midrule
        CALLHOME1 & $>$\SI{0}{\second}    & \SI{3}{\hour}    & 0.090 & 0.141 & \textcolor[gray]{0.7}{1.000} & \textcolor[gray]{0.7}{1.000} & 0.968 & 0.962 & 0.956 & 0.930 \\
        CALLHOME2 & $>$\SI{0}{\second}    & \SI{3}{\hour}    & 0.098 & 0.131 & 0.968 & 0.962 & \textcolor[gray]{0.7}{1.000} & \textcolor[gray]{0.7}{1.000} & 0.951 & 0.939 \\
        CSJ       & $>$\SI{0}{\second}    & \SI{12}{\hour}   & 0.168 & 0.201 & 0.956 & 0.930 & 0.951 & 0.939 & \textcolor[gray]{0.7}{1.000} & \textcolor[gray]{0.7}{1.000}  \\
        \midrule
        Concat-and-sum ($\beta=2$) & 
        $>$\SI{1.5}{\second}  & \SI{2461}{\hour} & 0.212 & 0.341 & 0.425 & 0.363 & 0.416 & 0.350 & 0.428 & 0.355 \\
        Concat-and-sum ($\beta=7$) & 
        $>$\SI{1.5}{\second}  & \SI{2428}{\hour} & 0.550 & 0.137 & 0.018 & 0.369 & 0.018 & 0.357 & 0.018 & 0.362 \\
        Ours (Random selection)                     & 
        $>$\SI{1.5}{\second}  & \SI{2459}{\hour} & 0.075 & 0.141 & 0.954 & 0.859 & 0.964 & 0.828 & 0.936 & 0.822 \\
        Ours (Markov selection)             & 
        $>$\SI{1.5}{\second}  & \SI{2473}{\hour} & 0.074 & 0.144 & 0.954 & 0.861 & 0.966 & 0.829 & 0.937 & 0.824 \\
        \midrule
        Concat-and-sum ($\beta=2$) & 
        $>$\SI{0}{\second}    & \SI{2401}{\hour} & 0.305 & 0.262 & 0.437 & 0.611 & 0.428 & 0.590 & 0.440 & 0.599 \\
        Concat-and-sum ($\beta=7$) & 
        $>$\SI{0}{\second}    & \SI{2393}{\hour} & 0.661 & 0.093 & 0.019 & 0.630 & 0.018 & 0.609 & 0.019 & 0.618 \\
        Ours (Random selection)                     & 
        $>$\SI{0}{\second}    & \SI{2452}{\hour} & 0.107 & 0.119 & 0.954 & 0.862 & 0.965 & 0.891 & 0.936 & 0.865 \\
        Ours (Markov selection)             & 
        $>$\SI{0}{\second}    & \SI{2458}{\hour} & 0.106 & 0.121 & 0.954 & 0.861 & 0.966 & 0.890 & 0.937 & 0.864 \\
        \bottomrule
    \end{tabular}%
    }
\end{table*}
The statistics of the real and simulated conversation data are shown in \autoref{tab:data_similarity}.
We report the silence and overlap ratios in addition to the silence and overlap similarities of each dataset against the CALLHOME1, CALLHOME2, and CSJ datasets.
First, the silence similarity and overlap similarity between the three real datasets are both high ($>0.9$), while the silence and overlap ratios are bit far apart between CALLHOME1/2 and CSJ.
This suggests that these similarities are more important in evaluating the naturalness of conversational datasets than the silence or overlap ratios themselves.

With the conventional concat-and-sum approach, the silence and overlap similarities are low ($<0.5$) when $\beta=2$.
$\beta=7$ brought the overlap ratio closer to the real datasets, but resulted in only a slight improvement in overlap similarity with severe degradation in silence similarity.
Using the utterances of less than $\SI{1.5}{\second}$ improves the overlap similarities to about 0.6, but they are still not as good as those between the real datasets.

With the proposed simulation method using the parameters extracted from the CALLHOME1 dataset, we obtained about 0.95 silence similarity and 0.86 overlap similarity with CALLHOME1.
Using the utterances of less than $\SI{1.5}{\second}$ did not improve the similarities with CALLHOME1, but it did improve the overlap similarities with CALLHOME2 and CSJ.
Note that the type of utterance transition generation (random selection or Markov selection) did not affect the similarities very much.
The reason is that silence or overlap similarity only consider the distribution of their duration, not their order.

\subsection{Diarization performance}
\begin{table*}[t]
    \caption{\textit{DERs (\%) on CALLHOME2 and CSJ.}}
    \label{tab:DER}
    \centering
    \resizebox{\linewidth}{!}{%
        \begin{tabular}{@{}lrrrrrrrrr@{}}
            \toprule
            &&\multicolumn{4}{c}{CALLHOME2}&\multicolumn{4}{c}{CSJ}\\\cmidrule(lr){3-6}\cmidrule(l){7-10}
             &  &
            \multicolumn{2}{c}{Transformer} & \multicolumn{2}{c}{Conformer}&
            \multicolumn{2}{c}{Transformer} & \multicolumn{2}{c}{Conformer}\\\cmidrule(lr){3-4}\cmidrule(lr){5-6}\cmidrule(lr){7-8}\cmidrule(l){9-10}
            Simulation method&Utt. length &
            w/o adapt & w/ adapt & w/o adapt & w/ adapt &
            w/o adapt & w/ adapt & w/o adapt & w/ adapt\\\midrule
            Concat-and-sum ($\beta=2$) & 
            $>$\SI{1.5}{\second} & 14.46 & 7.90  & 18.03 & 8.36 & 19.30 & 16.98 & 20.42 & 15.95 \\
            Concat-and-sum ($\beta=7$) & 
            $>$\SI{1.5}{\second} & 10.64 & 8.66 & 12.21 & 10.02 & 18.95 & 17.09 & 19.79 & 22.79 \\
            Ours (Random selection) & 
            $>$\SI{1.5}{\second} & 10.35 & 7.63 & 12.40 & 7.94 & 18.85 & 18.58 & 21.01 & 17.43 \\
            Ours (Markov selection)             & 
            $>$\SI{1.5}{\second} & 10.32  & 7.74 & 11.97 & 7.28& \textbf{16.99} & \textbf{14.36} & 20.30 & 16.23 \\
            \midrule
            Concat-and-sum ($\beta=2$) & 
            $>$\SI{0}{\second} & 12.26 & 7.80 &  11.25 & 7.53 & 19.41 & 17.00 & 18.86 & 16.55 \\
            Concat-and-sum ($\beta=7$) & 
            $>$\SI{0}{\second} & 10.11 & 8.61 & 10.70 & 8.93 & 19.51 & 17.92 & 19.37 & 21.15 \\
            Ours (Random selection) & 
            $>$\SI{0}{\second} & 10.39 & 7.83 &  9.74 & 7.32 & 17.96 & 17.71 & 17.52 & 14.50 \\
            Ours (Markov selection)      & 
            $>$\SI{0}{\second} & \textbf{9.65} & \textbf{7.53} & \textbf{9.65} & \textbf{7.18} & 17.85 & 15.26 & \textbf{17.09} & \textbf{12.41} \\\bottomrule
        \end{tabular}%
    }
\end{table*}

\autoref{tab:DER} shows the DERs of the Transformer-based and Conformer-based EEND-EDA on the real datasets: CALLHOME2 and CSJ.
Each model was evaluated with and without domain adaptation on CALLHOME1.
We observed that Conformer-based EEND-EDA trained on the dataset simulated with $>$\SI{0}{\second} utterances using the proposed method with Markov selection performed best, regardless of the adaptation.

In terms of the utterance length, Transformer-based EEND-EDA performed better than Conformer-based EEND-EDA when only the utterances longer than \SI{1.5}{\second} were used for simulation.
This indicates that Conformer, which takes temporal context into account, could only recognize speech segments lasting more than \SI{1.5}{\second} if only the utterances longer than \SI{1.5}{\second} were used for simulation.
In contrast, Conformer-based EEND-EDA was better than Transformer-based EEND-EDA when all the utterances were used for simulation, and in most cases the DERs were better than the cases where the utterances longer than \SI{1.5}{\second} were used.
From here, we focus on the results of Conformer-based EEND-EDA trained using the simulation datasets created with all the utterances ($>$\SI{0}{\second}).

In terms of the simulation methods, the datasets that were simulated with the proposed method contributed to improving the DERs.
Comparing the proposed method using random selection and using Markov selection, Markov selection was always better due to the improvement of the overlap similarity.
These results clearly show that the EEND performance can be improved by making the simulated mixtures more like real conversations, without using real data for pretraining as in \cite{liu2021end}.

Finally, we discuss the differences in EEND-EDA with and without domain adaptation.
Adjusting the value of $\beta$ of the concat-and-sum approach to obtain an overlap ratio similar to the real datasets helped improve DERs without adaptation ($\beta=2$ vs. $\beta=7$).
However, the DERs after adaptation were rather worse in the case of $\beta=7$ than in the case of $\beta=2$.
We presume this depends on whether the model overfits to the adaptation data (CALLHOME1) or generalizes well to the real data.
In \autoref{tab:data_similarity}, we can see that concat-and-sum with $\beta=7$ is closer to the real datasets than that with $\beta=2$ and thus was overfitted to CALLHOME1 during the adaptation phase; therefore, the DER on CALLHOME2, which has the same domain as CALLHOME1, was improved from \SI{10.70}{\percent} to \SI{8.93}{\percent}, but the DER on CSJ, which has a different domain, was degraded from \SI{19.37}{\percent} to \SI{21.15}{\percent}.
These results indicate that the improvement of the similarities between simulated and real datasets is the key to improving the DER on CSJ.
The datasets that are simulated using the proposed method, which are more similar to the real datasets, indeed contributed to improving the DERs, especially on CSJ: from \SI{17.52}{\percent} to \SI{14.50}{\percent} with random selection and from \SI{17.09}{\percent} to \SI{12.41}{\percent} with Markov selection.

\section{Conclusion}
In this paper, we proposed a method for simulating conversations that contain natural turn-taking.
We introduced four utterance transition types and sequentially arranged utterances in accordance with these types to simulate a long-form conversation.
The datasets that were simulated using the proposed method showed higher similarity with the real dataset than those simulated using a conventional method.
The experimental results showed that EEND-EDA pretrained using the datasets simulated by the proposed method contributed to improving DERs on both CALLHOME and CSJ.

We demonstrated the effectiveness of the proposed method under the two-speaker condition, but since \autoref{alg:mixture_simulation} is described in a way that does not limit the number of speakers, our future work will include evaluations in cases of more than two speakers.
In addition, since the diarization performance was improved by using Markov selection instead of random selection, further improvement can be expected by determining the utterance transition type according to a higher-order Markov chain.
Finally, the similarity metric used in this study does not capture the difference between the datasets created using random selection and Markov selection.
Constructing a new similarity metric that can express this difference is thus a possible avenue for future work.

\bibliographystyle{IEEEbib}
\bibliography{Odyssey2022_BibEntries}

\end{document}